\documentclass[preprint,aps]{revtex4}
\usepackage[dvips]{epsfig}
\begin{document}

\title{Phase resonances in obliquely illuminated \\compound gratings}
\author{Ricardo A. Depine$^{1}$, \'Angela N. Fantino$^{1}$, Susana I. Grosz$^{2}$ \\and Diana C. Skigin$^{1}$\\
{\em Grupo de Electromagnetismo Aplicado,}\\
{\em Departamento de F\'{\i}sica,}\\
{\em Facultad de Ciencias Exactas y Naturales, }\\
{\em Universidad de Buenos Aires, }\\
{\em Ciudad Universitaria, Pabell\'{o}n I, }\\
{\em C1428EHA Buenos Aires, Argentina}}
\date{July 2005}
\maketitle

\setcounter{footnote}{1} 
\footnotetext{Member of CONICET}
\setcounter{footnote}{2} 
\footnotetext{Ciclo B\'asico Com\'un, Universidad de Buenos Aires}

\baselineskip 4.5ex

\section*{Abstract}

The existence of phase resonances in obliquely illuminated, perfectly conducting compound gratings 
is investigated. The diffraction problem of a $p$-polarized plane wave impinging on the structure is solved 
using the modal approach. The results show that even under oblique illumination, where no
symmetry is imposed by the incident field, there are resonant wavelengths that are clearly
associated with a certain degree of symmetry in the phase distribution of the 
magnetic field inside the cavities. New configurations of this phase distribution take place,
that were not allowed under normal incidence conditions. It was found that the interior field is 
intensified in the resonances, and the specularly reflected efficiency is maximized. In particular,
this efficiency is optimized for Littrow mount.

\newpage

\section{Introduction}

It is well known that different kinds of resonances can be found in metallic corrugated structures:
surface plasmon polariton excitations (SPP), surface shape resonances (SSR) and phase resonances (PR).
SPPs are excited in infinite metallic gratings when illuminated by $p$-polarized light \cite{Agranovich}.
This excitation is accompanied by a 
significant power absorption \cite{Hessel, Hutley}, and consequently it
produces a sudden change in the efficiency curves of the reflected orders.  
For a given period and material of the grating, and for a fixed angle of
incidence, the excitation of a SPP is produced for a particular wavelength
at which one of the diffracted orders propagates parallel to the surface, and
therefore, the electric field near the surface is intensified.
This phenomenon is particularly important when the corrugations are shallow.
The SSRs, on the other hand, appear when the depth of the grooves is increased: 
the eigenmodes of each cavity can be
excited, producing interesting resonant effects such as field enhancement
inside the corrugations \cite{Valencia}-\cite{diana14}. Contrary to the
SPP excitations, these resonances are associated with the particular shape 
of each groove and can also be excited by $s$-polarized incident light \cite{Andrewartha}- 
\cite{Maradudin2}, but are independent of the period of the grating and the 
incidence angle. Experimental evidence of the SSR excitations was given
by L\'opez-Rios et al. \cite{Lopez-Rios}, for lamellar gratings. 

Another kind of resonances that might appear in structures with embedded cavities are 
the phase resonances. These resonances have been first reported in connection with structures 
comprising a finite number of cavities \cite{fikioris}-\cite{dv1}. These structures can be 
regarded as passive antennas, which exhibit superdirectivity \cite{bloch} for particular 
incidence conditions. When a bunch of cavities is illuminated by a $p$-polarized plane wave 
of a certain resonant wavelength, the far field pattern is narrowed while the field inside 
the cavities is enhanced and the phase difference between the magnetic field at adjacent grooves 
is $0$ or $\pi$ radians \cite{dv1}. This particular phase distribution, which is automatically 
generated by the resonant wavelength, produces a superdirective pattern \cite{Veremey2, Veremey1, dv1}.

Phase resonances are not allowed in infinite periodic simple gratings (gratings with a single groove 
in the period), since the pseudoperiodicity condition does not permit the field inside the
cavities to have different phases. However, PR in compound gratings under normal
illumination have already been reported \cite{sad1}. In this case, the specular efficiency is
maximized, the interior field is intensified, and the phase distribution of the field inside the
grooves takes forms analogous to those of finite gratings. The dependence of these resonances on
the different parameters of the grating was studied in \cite{sad2}. The existence 
of phase resonances in metallic gratings with ohmic losses have also been investigated recently  
\cite{sad3}. Other works on resonant excitations due to dual-period gratings include the paper by 
Hibbins et al. \cite{Hibbins},
where the authors study the excitation of surface plasmons in metallic structures in the microwave
region.

To deal with infinite metallic gratings of particular profiles (rectangular, triangular, semicircular), 
several modal approaches have been developed during the last four decades \cite{Andrewartha},\cite{Wirgin}, \cite{Jovicevic}-\cite{Li}. More recently, those modal methods have been generalized to arbitrary 
shapes of the corrugations, and numerical tools have been applied to enhance their performance 
\cite{Fox}-\cite{MerleElson}. To solve the diffraction problem from a perfectly conducting compound 
grating with rectangular grooves we applied the formulation proposed by Andrewartha {\em et. al} 
for simple gratings \cite{Andrewartha}, and extended it to compound gratings \cite{sad1}. 
This method is particularly
suitable for rectangular profiles, and enables us to analyze the electromagnetic response in terms
of eigenmodes, which is convenient for studying resonances.

In this paper we investigate the excitation of phase resonances in compound gratings under oblique
illumination. We show that the occurrence of such resonances is intimately connected with the
phase distribution of the magnetic field inside the cavities. Particular attention is paid to the
symmetry of the phase configurations that are automatically generated by the resonant
wavelengths.
The diffraction problem and the modal method are briefly described in Sec. 2.
Numerical results that evidence the existence of phase resonances for non-normal incidence are
given in Sec. 3, where we show efficiency curves as well as the amplitude and phase of the interior
field as a function of the wavelength for a fixed angle of incidence, and for the order -1 and -2
Littrow mounts. The most significant conclusions are summarized in Sec. 4.

\section{Configuration and method of resolution}

We consider a $p$-polarized plane wave of wavelength $\lambda$ that illuminates a 
perfectly conducting compound diffraction grating comprising several rectangular 
grooves in each period, as shown in Fig. 1. The structure parameters are the period
($d$), the number of grooves in each period ($N$), the width ($a$) and the depth ($h$)
of each groove, and the distance between grooves ($b$). The wave vector $\vec{k}$ of the
incident plane wave forms an angle $\theta_0$ with the $y$-axis. 

A comment on the validity of the ideal model of a perfectly conducting structure is pertinent
here. At a first sight, one should think that for studying resonances it is basic to consider 
a real metallic surface. This is true if we are interested in surface plasmon excitations, or
if we are in the region of the spectrum where these excitations are relevant. However, the purpose 
of this paper is to study the excitation of phase 
resonances, which are independent from surface plasmon excitations, and arise from a particular
distribution of the field inside the cavities. The resonant wavelength is not mainly related with the 
refraction index of the metal, but with the particular arrangement of the phase of the electromagnetic
field inside the rectangular corrugations. Besides, our study is meant to deal with large
wavelengths (of the order and above the microwave
region of the spectrum), where the conductivity of metals is very high, and they behave as perfect 
conductors. 

The method of resolution is the modal method, which consists in expanding the fields 
inside the grooves in their own eigenfunctions satisfying the boundary conditions
at the lateral walls and at the bottom of each cavity. A complete and detailed description
of the modal method applied to the diffraction problem from compound gratings can
be found in Ref. \cite{sad1}. This method is very suitable for the rectangular profile
of the cavities considered here. In this case, it takes a very simple form and it permits us to
study the dependence of the phase resonances by varying independently the geometrical
parameters of the groove. It has also been shown to be efficient for the calculation
of the electromagnetic response of very deep gratings \cite{diana1}.

For the sake of completeness, we summarize in what follows the modal method applied to 
the present grating. The essence of the method, is to 
distribute the space into three regions: 
above the structure, inside the grooves, and in the perfect conductor. The diffracted 
magnetic field in the upper region is expressed as Rayleigh expansions, the total magnetic 
field inside the cavities is expressed by modal expansions, and the field inside the perfect 
conductor is zero. Then, by enforcing the matching conditions at the horizontal boundaries 
between regions, one yields a system of equations for the unknown amplitudes of the field inside 
and above the grooves, which is solved by a standard numerical technique for 
matrix inversion.

\section{Results}

In this section we show numerical results that account for the existence of phase resonances
in infinite perfectly-conducting compound gratings, when illuminated by an obliquely incident 
plane wave. Phase resonances under normal illumination have already been investigated for 
different numbers of grooves in the period \cite{sad1}, and for varying geometrical parameters 
of the structure \cite{sad2}. It was found that this kind of resonances appears for certain
wavelengths that maximize the specularly reflected efficiency as well as the electromagnetic field 
inside the cavities, and, at the same time, generates a particular distribution of the phase 
difference between the magnetic field at adjacent grooves, which is of either $\pi$ or $0$ 
radians \cite{sad1}. 

In what follows we analyze the response of gratings with $N=3$ and $N=5$ in three different 
situations: $\theta_0=50^\circ$, order -1 Littrow mount, and order -2 Littrow mount. 
The existence of phase resonances at oblique illumination is manifested in Fig. 2, where we
plot the specular efficiency (2a) and the field amplitude (2b) as a function of the 
wavelength-to-depth ratio $\lambda/h$, for normal incidence and for 
$\theta_0=50^\circ$. The phase difference between the fundamental 
amplitudes of the magnetic field at the external grooves for $\theta_0=50^\circ$ is plot in Fig. 2c. This 
angle was arbitrarily chosen to illustrate the phenomenon at oblique incidence, since
the response of the structure is similar for any $\theta_0 \neq 0^\circ$. 
The parameters of the grating are $N=3$, $d/h=5.5$, $a/h=0.3$, $b/h=0.1$. It can be observed 
in Fig. 2(a) that for $\theta_0=50^\circ$ the specular efficiency has two peaks in the range of
$\lambda$ considered, one at nearly the same wavelength as for normal incidence 
($\lambda/h \approx 4.53$), and another one around $\lambda/h \approx 5.05$. 
This confirms the existence of phase resonances under oblique illumination. Moreover, there
is a new resonant wavelength that is not present in the normal incidence case. The amplitude 
of the fundamental mode of the magnetic field inside the different cavities, which
in this case is nearly equivalent to the total field amplitude in each groove, is plot in Fig. 2(b). 
It can be observed that for normal
illumination, the field is maximized at the resonant wavelength, as it was already shown 
in Ref. \cite{sad1}. On the other hand, for $\theta_0=50^\circ$ the amplitudes of the external grooves are 
coincident for $\lambda/h=4.53$, and are very similar in the interval $\lambda/h \in [4.6,5.05]$
(Fig. 2b, dotted and dashed-dotted curves). It can be observed in Fig. 2c that the phase difference
between the fundamental mode amplitudes at the external grooves is zero for $\lambda/h=4.53$ and it is $\pi$ 
radians for $\lambda/h=4.6$ and 5.08. The resonance at $\lambda/h=4.53$ can be associated with the $\pi$ resonance in the normal incidence case, in which the phase difference between the external grooves is 
zero, and it is $\pi$ between each one of these grooves and the central one. The second resonance, however,
appears at $\lambda/h=5.05$, which does not correspond to any of the values where the phase difference is
exactly $\pi$. At $\lambda/h =4.6$ the phases of the external cavities are not symmetric with respect to that of the central one, and since the central amplitude is comparable to 
that of the external grooves, it prevents the generation of an effective $\pi$ radians difference
necessary to have a phase resonance. On the other hand, for $\lambda/h=5.08$ the phase difference is
exactly $\pi$, but although the central groove amplitude is low, for this wavelength the amplitudes 
of the external grooves are no longer equal, and then the symmetry requirement is not fulfilled.
In summary, the value at which all three conditions (symmetry in the amplitude distribution, $\pi$ phase difference between the external cavities and low amplitude of the central one) 
seem to be better achieved is $\lambda/h=5.05$. This suggests that the generation of phase resonances 
is connected with the phase difference between grooves but also with a certain kind of symmetry in the 
field distribution. It is important to remark that both requirements are concerned with symmetries that
are automatically generated in the structure, even under non-symmetric incidence conditions.

Taking into account that the phase is almost constant within each groove, we can represent the amplitude 
and phase of the magnetic field at the central position ($x_j=a/2+j*(a+b)$, $j=0,...,N-1$) of the bottom 
($y=-h$) of each groove as a phasor, for the resonant wavelengths. 
The phase distribution for the first peak in Fig. 2a is nearly the same as that of the $\pi$-mode at normal incidence
\cite{dv1,sad1}, i.e., the phases of the external grooves are equal, and differ in $\approx \pi$ radians 
from that of the central groove. On the other hand, the phase distribution
for the second resonance (second peak in Fig. 2a) is different: the phase difference between the fields of the external grooves
is almost $\pi$ radians, and the central groove has an intermediate phase which differs in $\approx \pi/2$ 
radians from that of the external grooves. Besides, the central field amplitude is negligible. 
This is a new situation that was
forbidden for normal incidence, and is generated by a resonant wavelength in oblique incidence.

In Fig. 3 we study the case of the order -1 Littrow mount, and compare the results obtained for the same 
grating considered in Figure 2 and for a grating with two grooves per period (N=2), with the same width $a$
but separated a distance $b^\prime/h=(2b+a)/h=0.5$. 
The numerical results suggest that the new resonant effect which appears under oblique incidence is improved 
in the -1 Littrow case. The specular efficiency for N=3
is optimized for two values of the wavelength (Fig. 3(a)). At these values the field inside 
the external grooves has nearly the same magnitude (Fig. 3(b)), although this symmetry is not imposed by the 
incidence conditions. The phase configurations for the first and the second peak are very similar to 
those corresponding to $\theta_0=50^\circ$. Notice that even though for
the first peak the central groove has the maximum amplitude, for the second peak the field amplitude 
at the central groove is significantly lower than that of the other grooves. This seems to be a
condition necessary to generate this second resonance: there is a phase difference of nearly $\pi$
radians between the fundamental modes of the external grooves (see Fig. 3c), and the central one 
does not play any role. Actually, if we compare the curves for N=2 and N=3 in Figs. 3a and 3b, we observe that the first resonance is not present for N=2 but the second is at
a close wavelength. This behavior can be explained as follows. To generate the first peak, a phase 
difference between the modal amplitudes at adjacent grooves of $\pi$ radians is needed. For N=3 the central 
groove has a significant amplitude and its phase is nearly opposite to the phases of the external grooves,
thus allowing the existence of the resonance. On the other hand, for N=2 the phase difference between
the grooves is close to zero, what inhibits the generation of the resonance (see Fig. 3c). For the second 
peak the
situation is different, since at this resonant wavelength the field amplitude in the central groove 
in the N=3 case is too low making this groove irrelevant for the generation of the resonance, and the 
$\pi$ phase difference needed is found between the external grooves. Then, replacing the central groove
by full metal does not change significantly the field distribution and the second resonance is also 
present in this case.

The order -2 Littrow mount is analyzed in Figs. 4 and 5. Note that the propagation direction of the 
diffracted orders is the same here as in normal incidence. Therefore, this case presents particular interest since, 
kinematically, it looks equivalent to the normal incidence situation: 
the order -1 propagates normally to the structure and the orders 0 and -2 propagate along symmetric 
directions with respect to the mean normal. In this case, the first resonant peak coincides 
with that of normal incidence,
and the second maximum is very wide and cannot be regarded as a resonance (see Fig. 4(a)). In fact, there
is no intensification of the interior field at this wavelength (Fig. 4(b)). In Fig. 5 we show the dependence 
on the angle of incidence of the specular efficiency (Fig. 5(a)), magnitude (Fig. 5(b)) and phase difference 
(Fig. 5(c)) of the magnetic field, for the resonant wavelength corresponding to normal incidence and to order -2 Littrow mount ($\lambda/h \approx 4.5302$). As expected, the specular efficiency has two maxima: one at $\theta_0=0^\circ$
and another at $\theta_0=55.45^\circ$ which corresponds to -2 Littrow mount, and a minimum at 
$\theta_0=24.32^\circ$, which corresponds to -1 Littrow mount. It can be observed in Fig. 5(c)
that the phase difference between the magnetic field at adjacent grooves for $\theta_0=55.45^\circ$
and for $\theta_0=24.32^\circ$ is equal, although the field magnitudes at the external grooves are 
equal only for -2 Littrow mount (Fig. 5(b)). 
This result suggests that the -2 Littrow situation optimizes the resonance, since it automatically generates a highly symmetric field distribution inside the grooves, resembling the high symmetry 
previously found for normal incidence. 

In Figures 6-7 we investigate a grating with 5 grooves in the period, in the -1 order Littrow mount.
It is important to recall that in the normal incidence case, for an increasing number of cavities in the 
period, the number of resonances is also increased \cite{sad1}. In the case of oblique incidence,
this behavior is also found, as it is shown in Fig. 6 for a grating with the same parameters as those 
of the previous figures, except for $N=5$. Comparing the specular efficiency curve with that corresponding 
to $N=3$ (Fig. 3(a)), we notice that the resonance at $\lambda/h=4.539$ is now split into two
peaks at $\lambda/h=4.475$ and $\lambda/h=4.575$ (Fig. 6). The distribution of the amplitude of the 
magnetic field in the vicinity of the grooves for the resonant wavelengths, is illustrated as contour plots in 
Fig. 7. The black zones represent the most intense fields, whereas the white zones represent the weakest
fields (the perfectly conducting region is light gray). The grey scale was chosen so as to make it easy to
visualize the field structure in each case, and therefore it is not maintained in all three figures. For
instance, the maximum value in Fig. 7(a) is about two times that of Fig. 7(b), and about five times
that of Fig. 7(c). 

For the first peak in Fig. 6 ($\lambda/h = 4.475$) there is a clear intensification of the field inside the grooves, the magnetic field being maximum at the central groove. The phase distribution corresponds to the 
$\pi$-mode
\cite{sad1}, i.e., phase differences of $\pi$ between adjacent grooves (Fig. 7(a)). 
For the
second resonance ($\lambda/h = 4.575$, Fig. 7(b)), there is also an enhancement of the field
within the cavities, even though it is not as strong as for the first peak, and the amplitude at the
central groove is very low ($\approx 1/25$ of the maximum amplitude inside the cavities). This second resonance 
is similar to the second peak for $N=3$ (Fig. 3), which was forbidden for normal incidence. The phase
of the central groove forms an angle of $\approx \pi/2$ with the phases of its adjacent grooves, which have
opposite phases; the external grooves have opposite phases too, which are also opposite to their 
corresponding neighbors.  
Finally, the intensification of the third peak ($\lambda/h = 4.89$) is not as 
important as in the previous two cases, and the distribution of phases is also different (Fig. 7(c)).
This situation seems to be close to the second resonance of normal incidence (see Fig. 5(b) in Ref.
\cite{sad1}). It is important to remark that for $\lambda$ different from the resonant values, the
interior field is significantly lower. For an increasing number of grooves in the period, new splittings
of the resonances were found, which give place to new resonant situations not allowed under normal incidence 
(not shown). 

The results show that phase resonances appear in perfectly conducting compound gratings 
not only under normal incidence, which is a highly symmetric situation, but also under oblique illumination.
It is important to remark that even though the generation of phase resonances is strongly connected with 
symmetries in the magnetic field distribution inside the cavities (amplitude and phase), these requirements 
are automatically fulfilled for particular wavelengths, even when the symmetry is not imposed by external conditions.  
   
\section{Summary and conclusions}

The existence of phase resonances under oblique illumination has been studied. For this purpose,
we investigated the response of a rectangular perfectly-conducting compound grating 
as a function of the wavelength (for a fixed angle of incidence), of the angle of
incidence (for a particular resonant wavelength) and for Littrow mount.
New resonant wavelengths, not allowed for normal incidence, were found. The magnitude and phase 
distributions of the magnetic field at these wavelengths exhibit a high degree of symmetry. 
The magnetic field is intensified inside the grooves, and the specularly reflected efficiency is 
optimized in Littrow mount. 
The physical mechanism that automatically generates these particular field configurations is not 
fully understood yet, but we are working towards this objective. We are also investigating on
the exploitation and optimization of this phenomenon for its use in optical devices. 

\section*{Acknowledgments}

The authors gratefully acknowledge partial support from Consejo Nacional de
Investigaciones Cient\'{\i}ficas y T\'ecnicas (CONICET), Universidad de Buenos Aires (UBA) 
and Agencia Nacional de Promoci\'on Cient\'{\i}fica y Tecnol\'ogica (ANPCYT-BID
802/OC-AR03-04457).

\newpage
\thispagestyle{empty}

\begin{figure}[h]
\includegraphics[width=10cm]{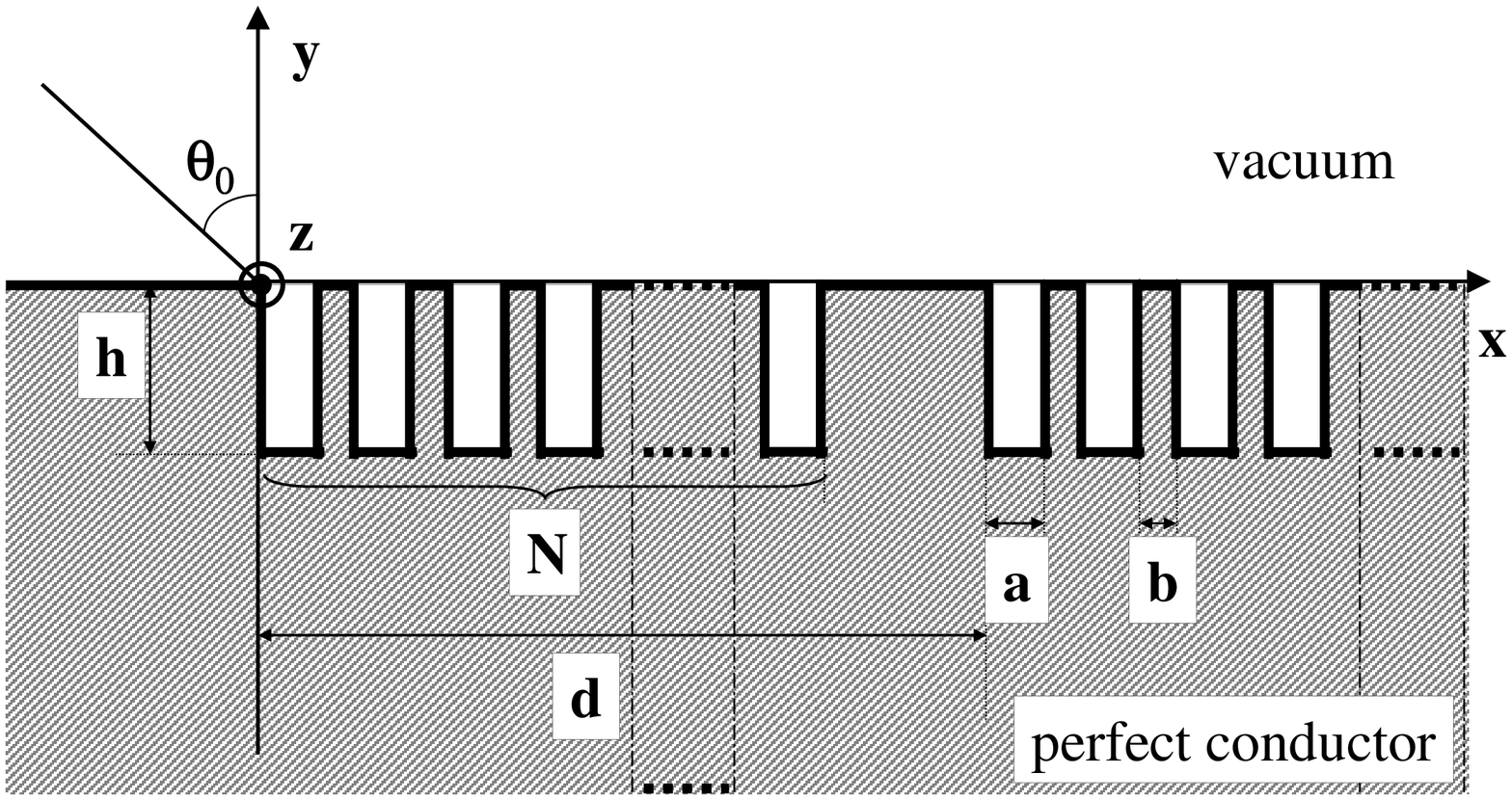}
\caption{Configuration of the problem.}
\end{figure}

\newpage
\thispagestyle{empty}

\begin{figure}[h]
\begin{tabular}{c}
\includegraphics[width=8cm]{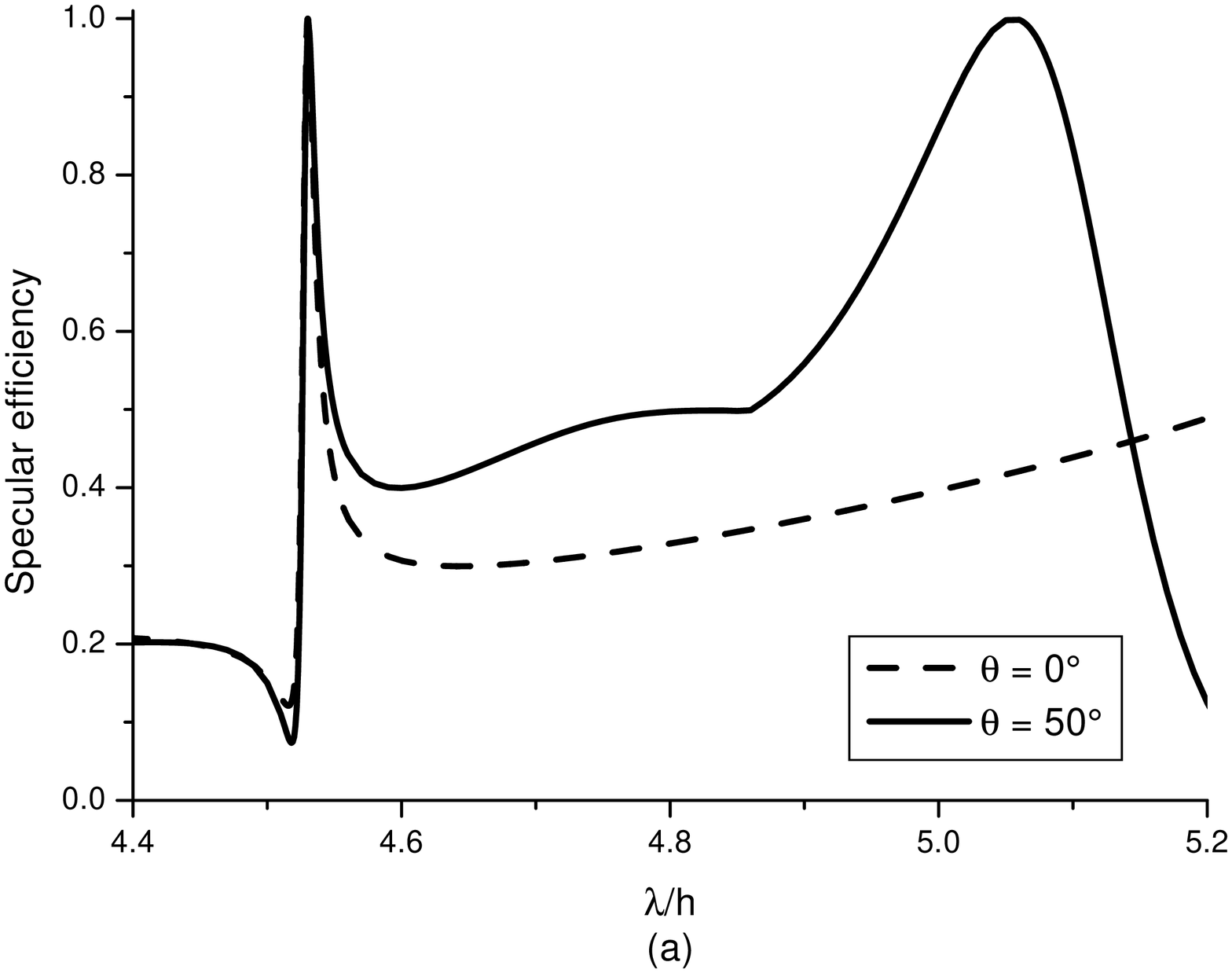} \\ %\hspace{0.3cm}
\hspace{-0.5cm}\includegraphics[width=8cm]{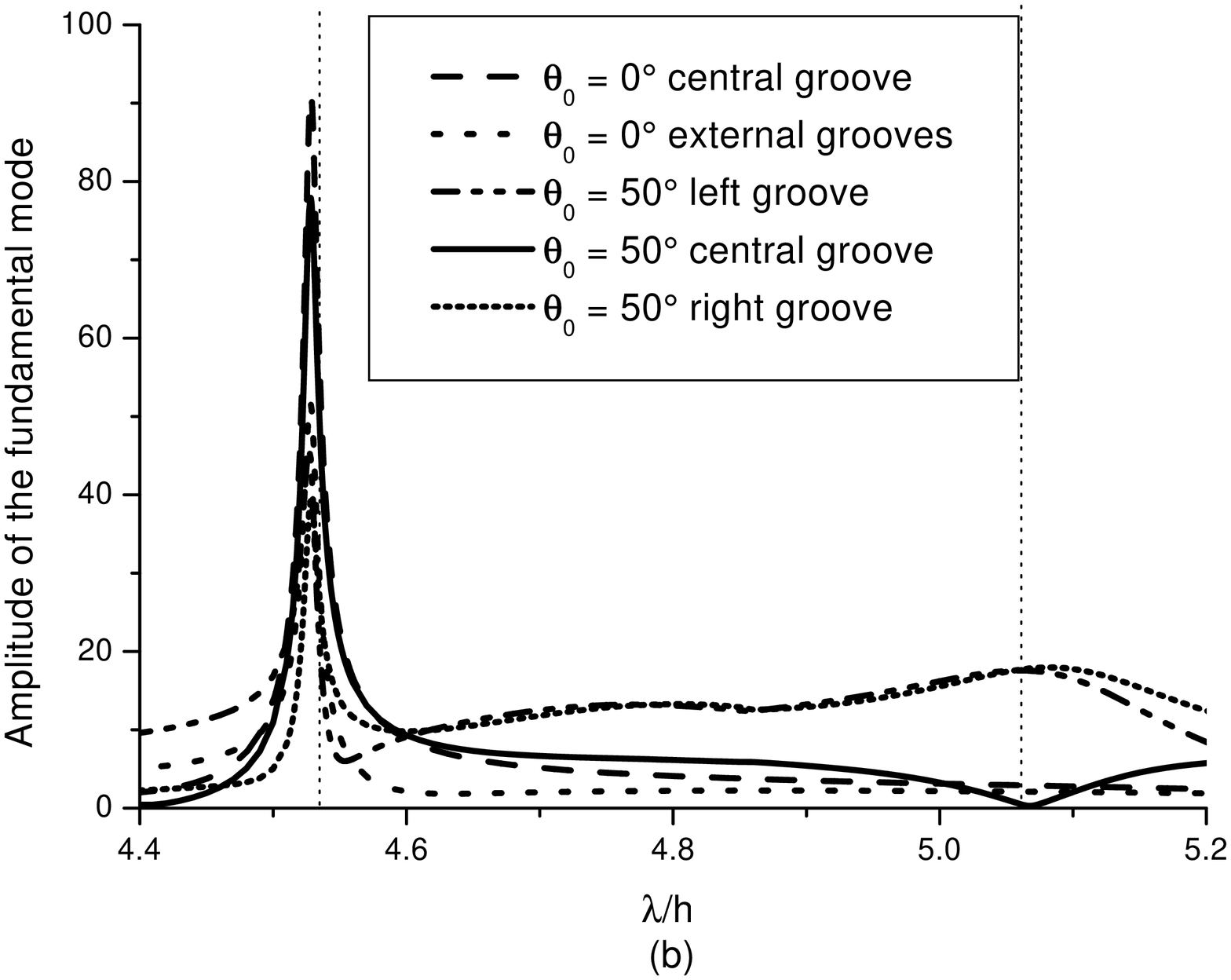} 
\includegraphics[width=8cm]{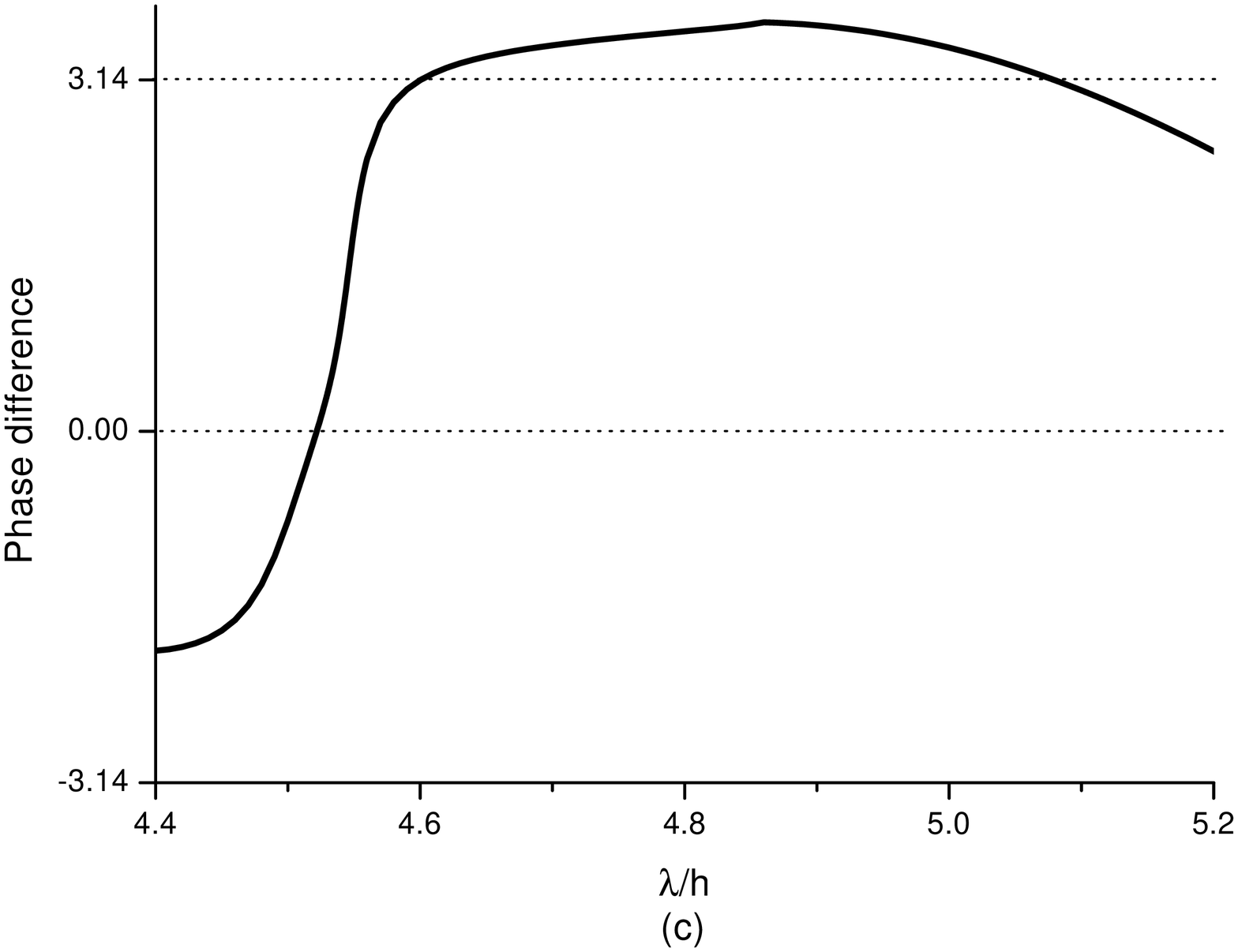} %\hspace{0.3cm}
\end{tabular}
\caption{(a) Specular efficiency versus $\lambda/h$ for $\theta_0=0^\circ$ and $\theta_0=50^\circ$; (b) Amplitude of the fundamental mode of the magnetic field inside the cavities versus $\lambda/h$ for $\theta_0=0^\circ$ and $\theta_0=50^\circ$; 
(c) Phase difference of the fundamental modes inside the cavities versus $\lambda/h$, for $\theta_0=50^\circ$. The parameters of the grating are 
$N=3$, $d/h=5.5$, $a/h=0.3$ and $b/h=0.1$.}
\end{figure}

\newpage
\thispagestyle{empty}

\begin{figure}[h]
\begin{tabular}{c}
\includegraphics[width=8cm]{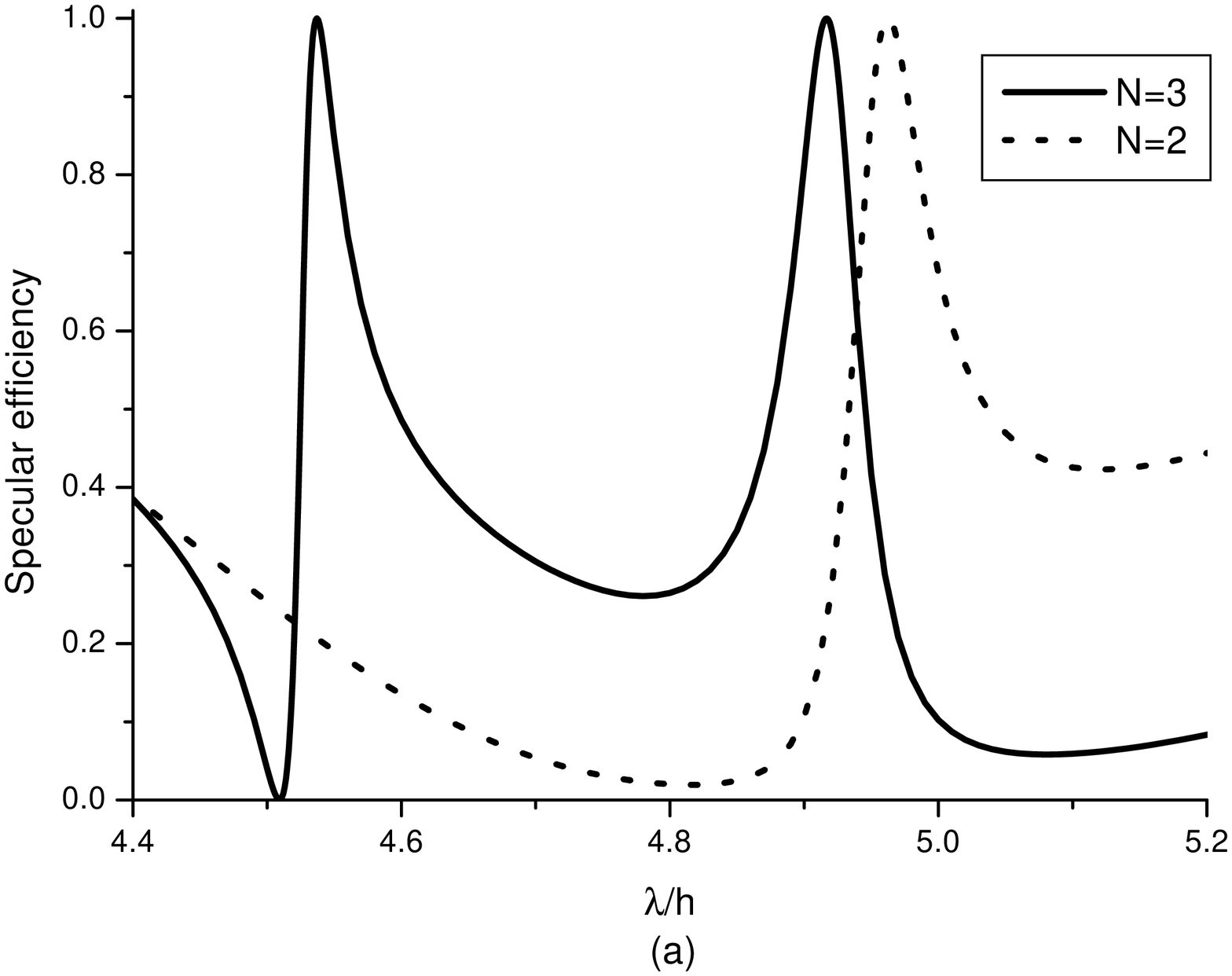} \\ %\hspace{0.3cm}
\hspace{-0.5cm} \includegraphics[width=8cm]{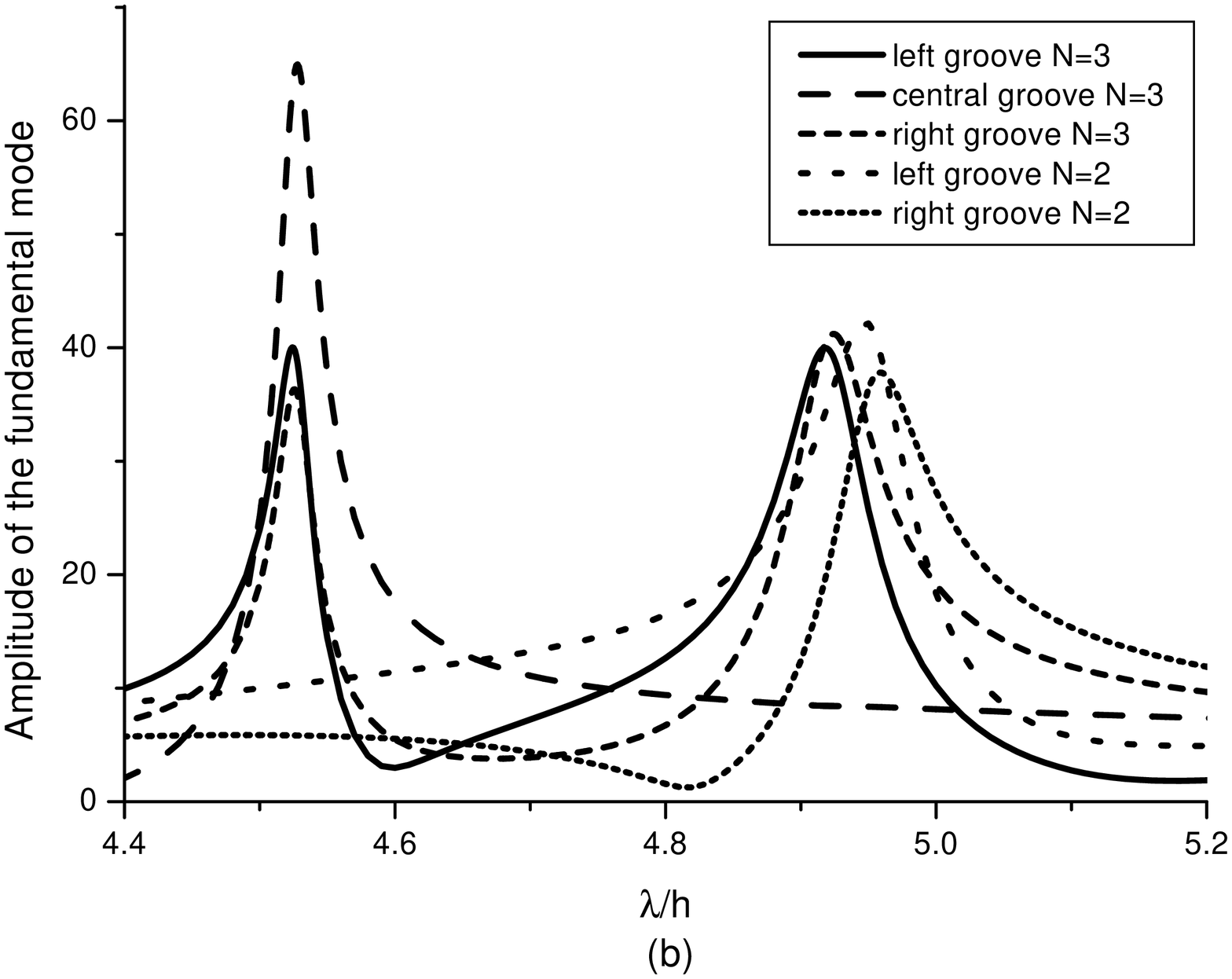} 
\includegraphics[width=8cm]{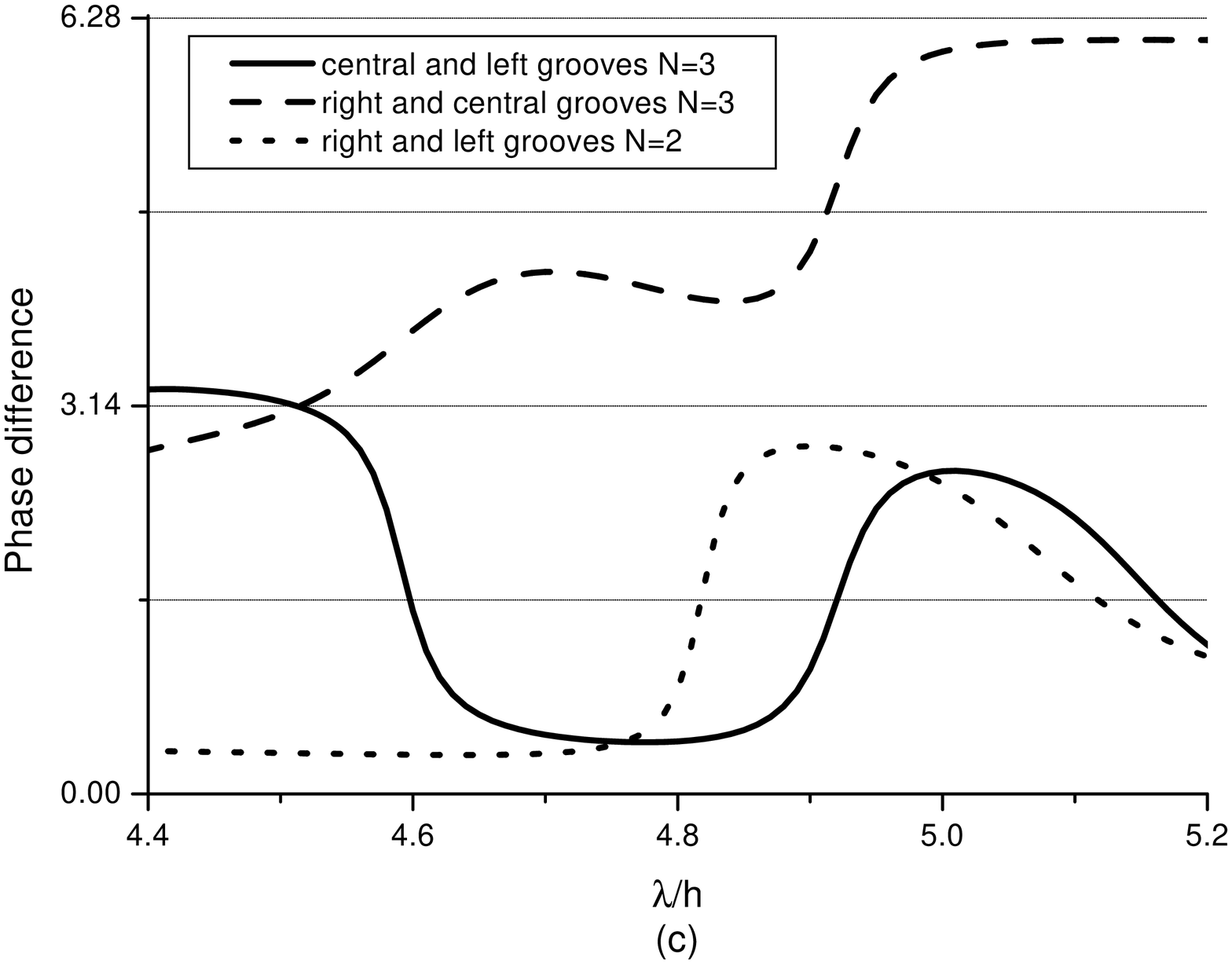} %\hspace{0.3cm}
\end{tabular}
\caption{(a) Specular efficiency versus $\lambda/h$ for the grating of Fig. 2 and for a grating with
N=2, $a/h=0.3$ and $b/h=0.5$, for the order -1 Littrow mount; (b) Amplitude of
the fundamental mode of the magnetic field inside the cavities versus $\lambda/h$, for 
the cases considered in (a); (c) Phase difference between the fundamental modes inside the cavities versus $\lambda/h$, for the cases considered in (a).}
\end{figure}

\newpage
\thispagestyle{empty}

\begin{figure}[ht]
\begin{tabular}{c}
\includegraphics[width=8cm]{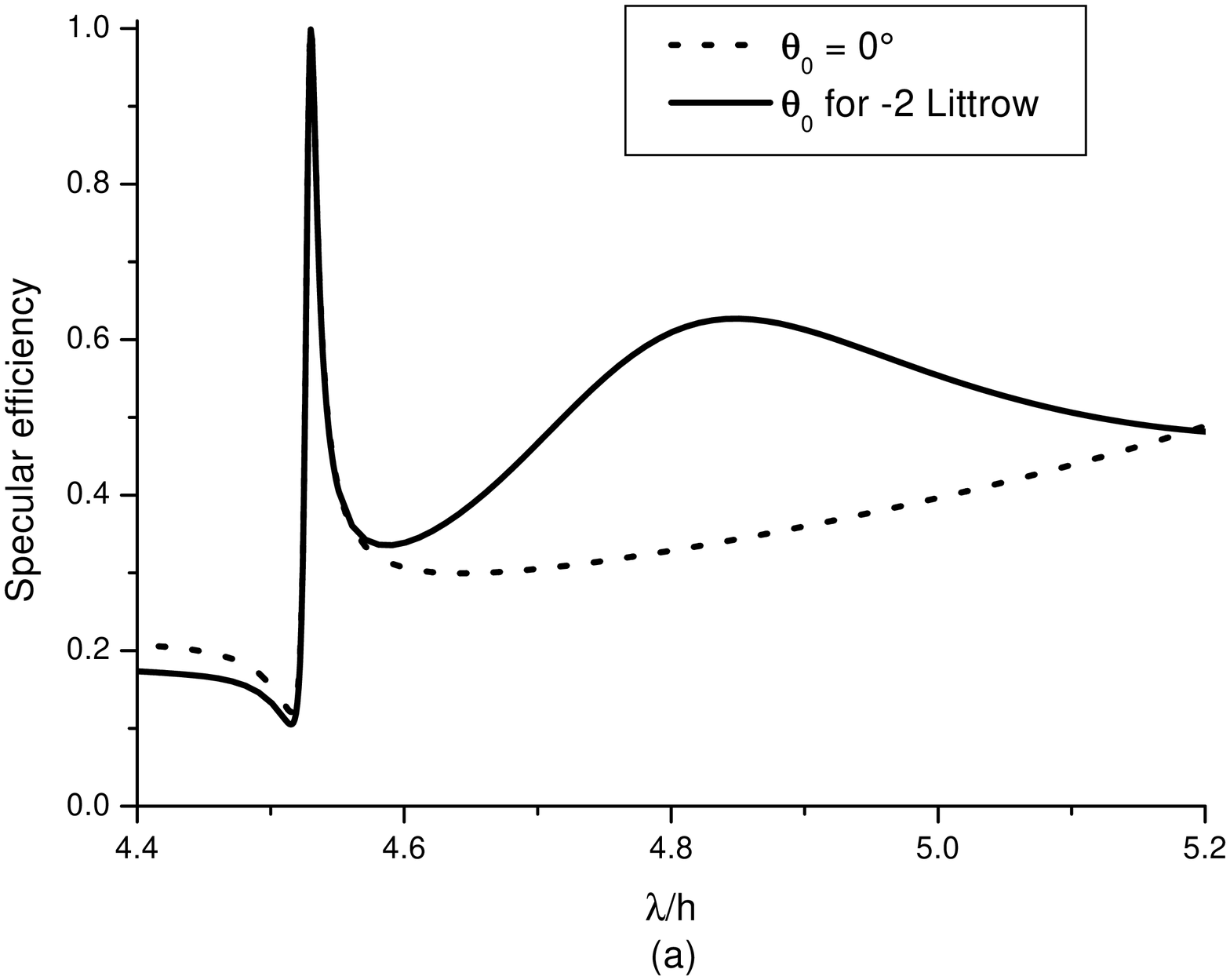}  %\hspace{0.3cm}
\hspace{-0.5cm} \includegraphics[width=8cm]{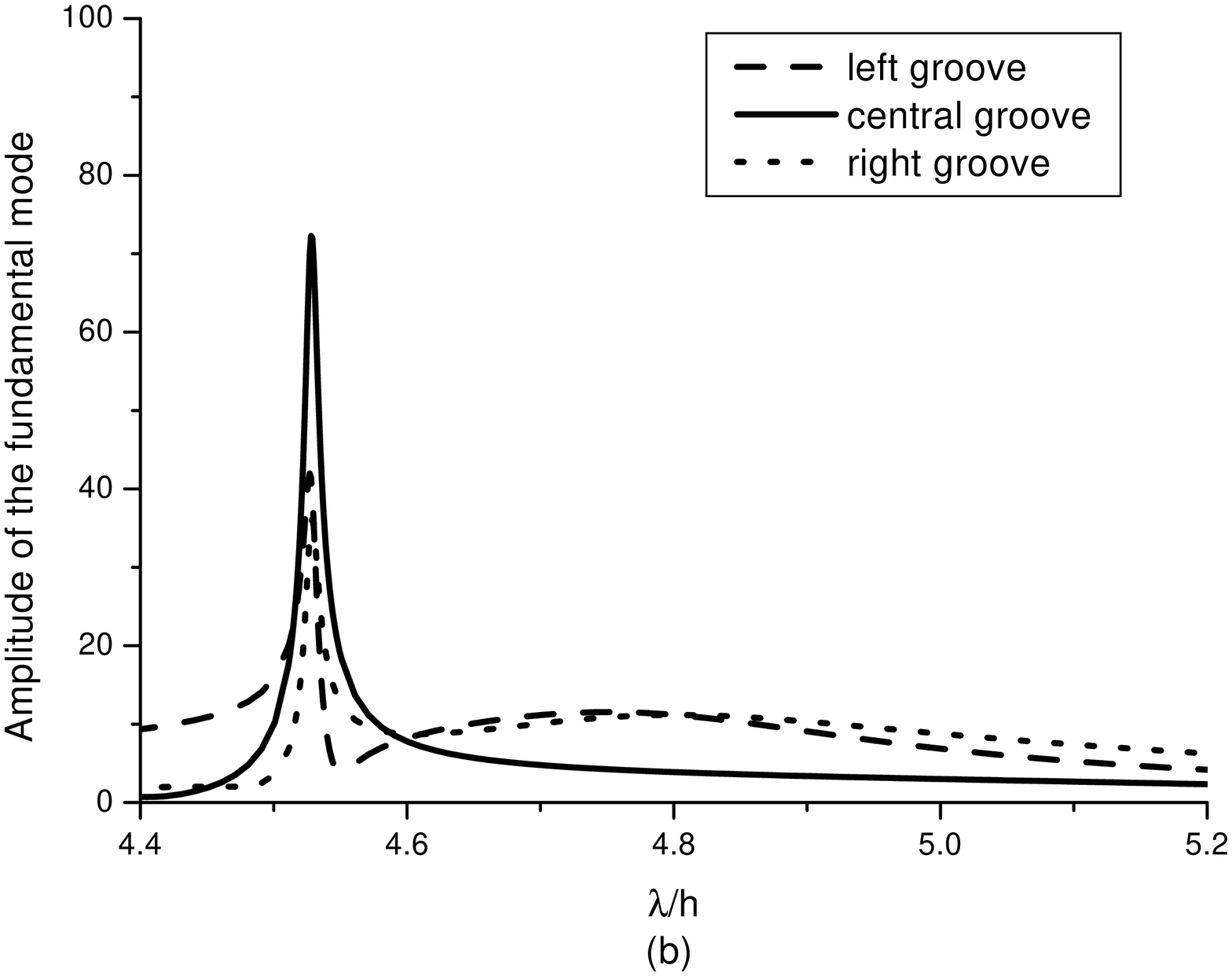} 
\end{tabular}
\caption{(a) Specular efficiency versus $\lambda/h$ for the grating of Fig. 2, for normal 
incidence and for the order -2 Littrow mount; (b) Amplitude of
the fundamental mode of the magnetic field inside the cavities versus $\lambda/h$, for 
the order -2 Littrow mount case.}
\end{figure}

\newpage
\thispagestyle{empty}

\begin{figure}[ht]
\begin{tabular}{c}
\includegraphics[width=8cm]{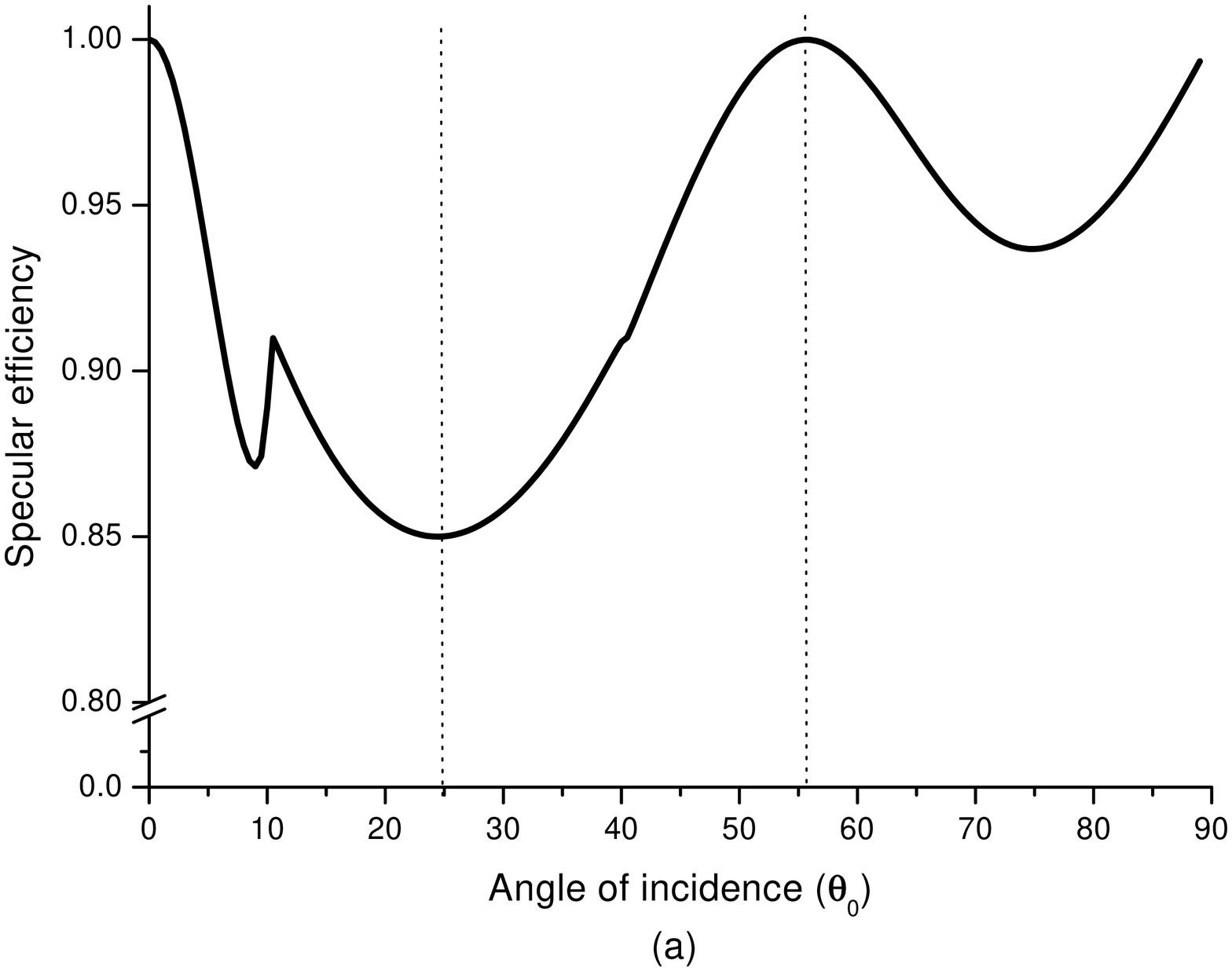} \\ %\hspace{0.3cm}
\hspace{-0.5cm} \includegraphics[width=8cm]{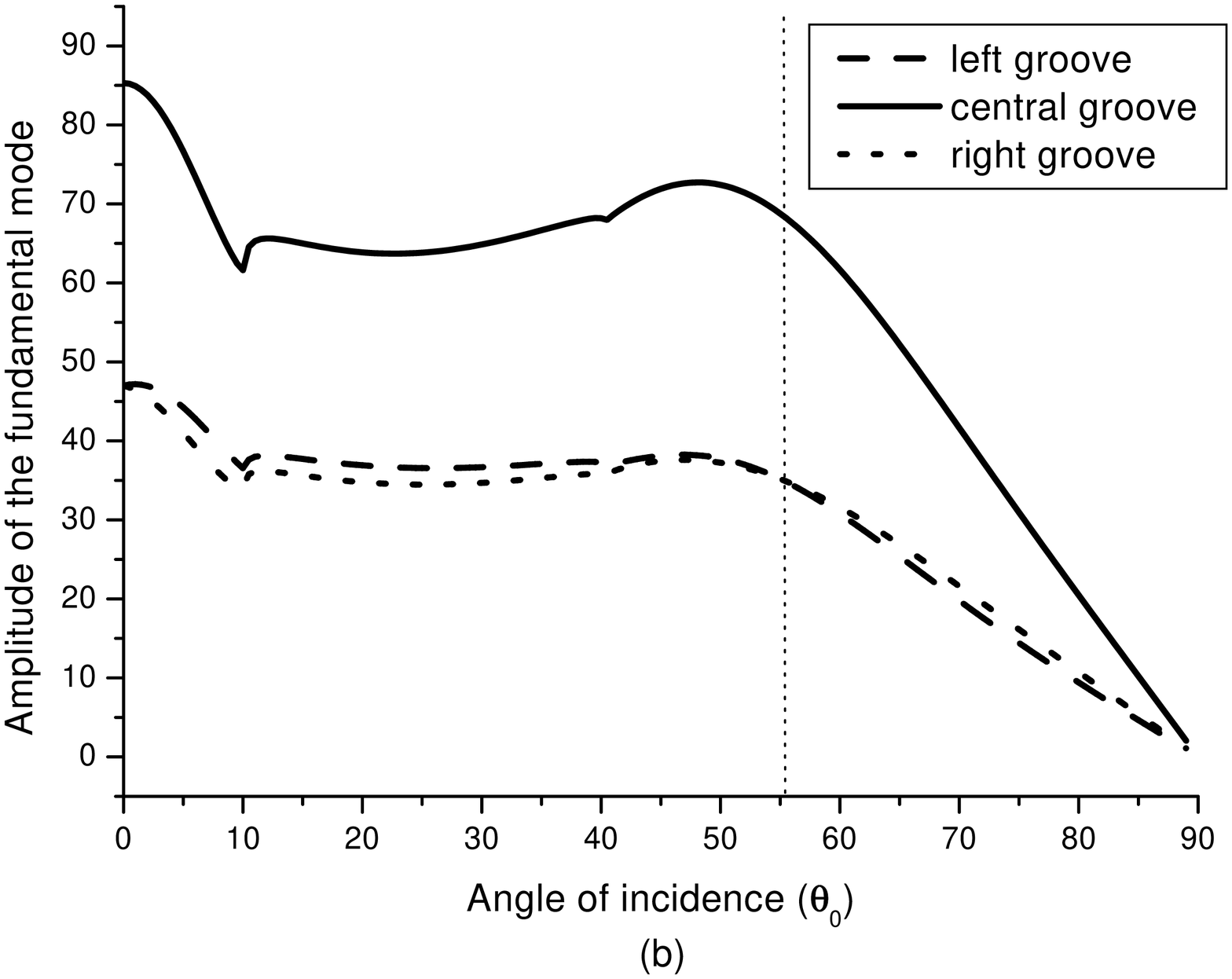} 
\includegraphics[width=8cm]{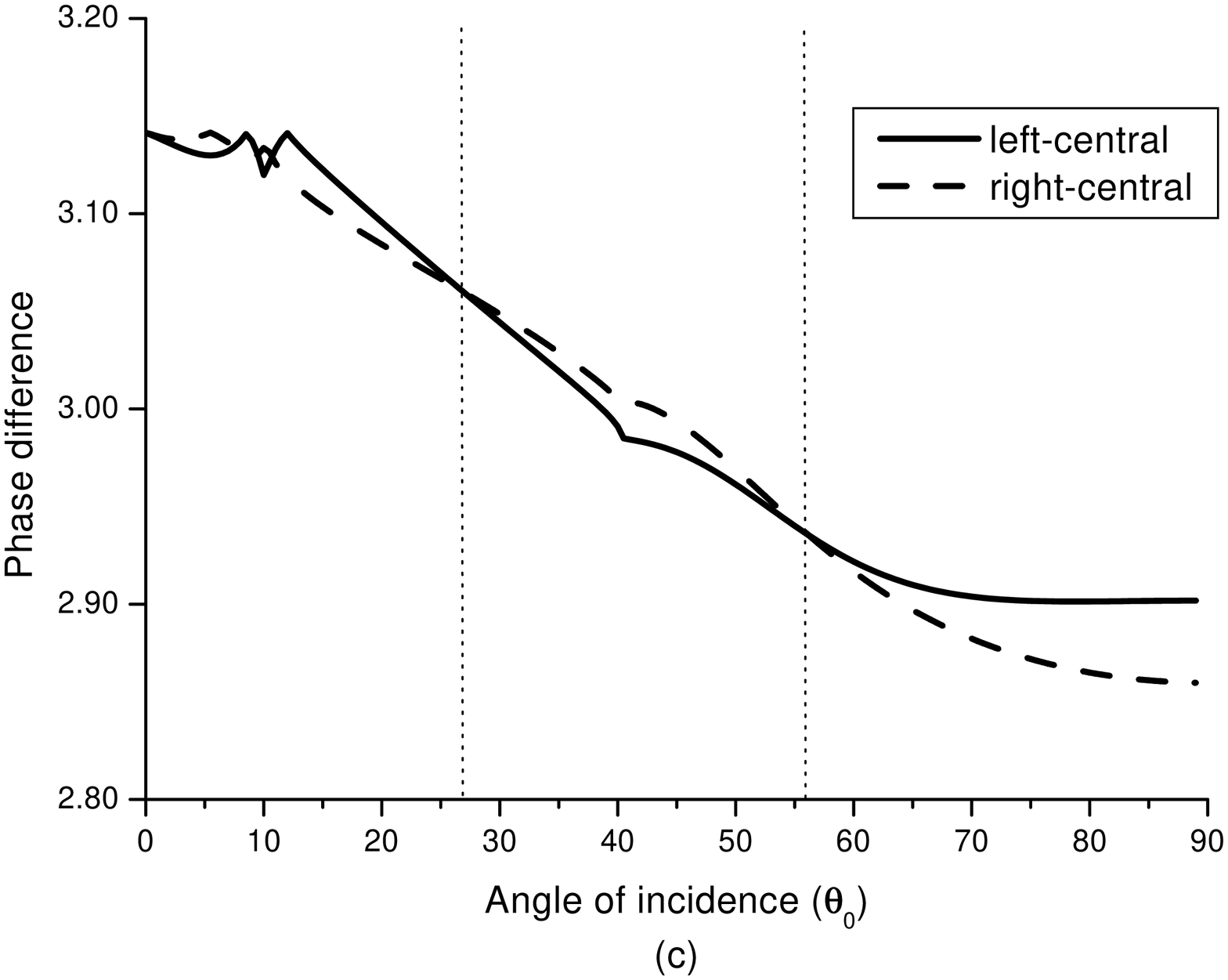} %\hspace{0.3cm}
\end{tabular}
\caption{(a) Specular efficiency; (b) Amplitude of the fundamental mode; (c) Phase difference 
between the fundamental mode of the magnetic in adjacent cavities, versus $\theta_0$, 
for the resonant wavelength corresponding to normal incidence and to the order -2
Littrow mount ($\lambda/h = 4.5302$). The parameters of the grating are the same as in Fig. 2.}
\end{figure}

\newpage
\thispagestyle{empty}

\begin{figure}[h]
\includegraphics[width=7cm]{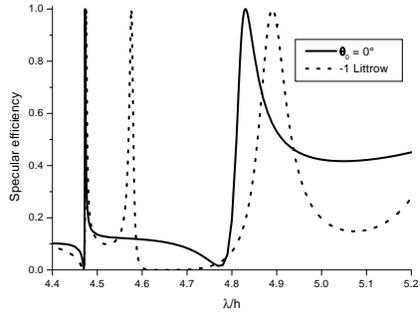}
\caption{Specular efficiency versus $\lambda/h$ for $\theta_0=0^\circ$ and for the order -1 Littrow 
mount, for the same grating of Fig. 2, except for $N=5$.}
\end{figure}

\newpage
\thispagestyle{empty}

\begin{figure}[ht]
\begin{tabular}{c}
\hspace{-1.5cm} \includegraphics[width=10cm]{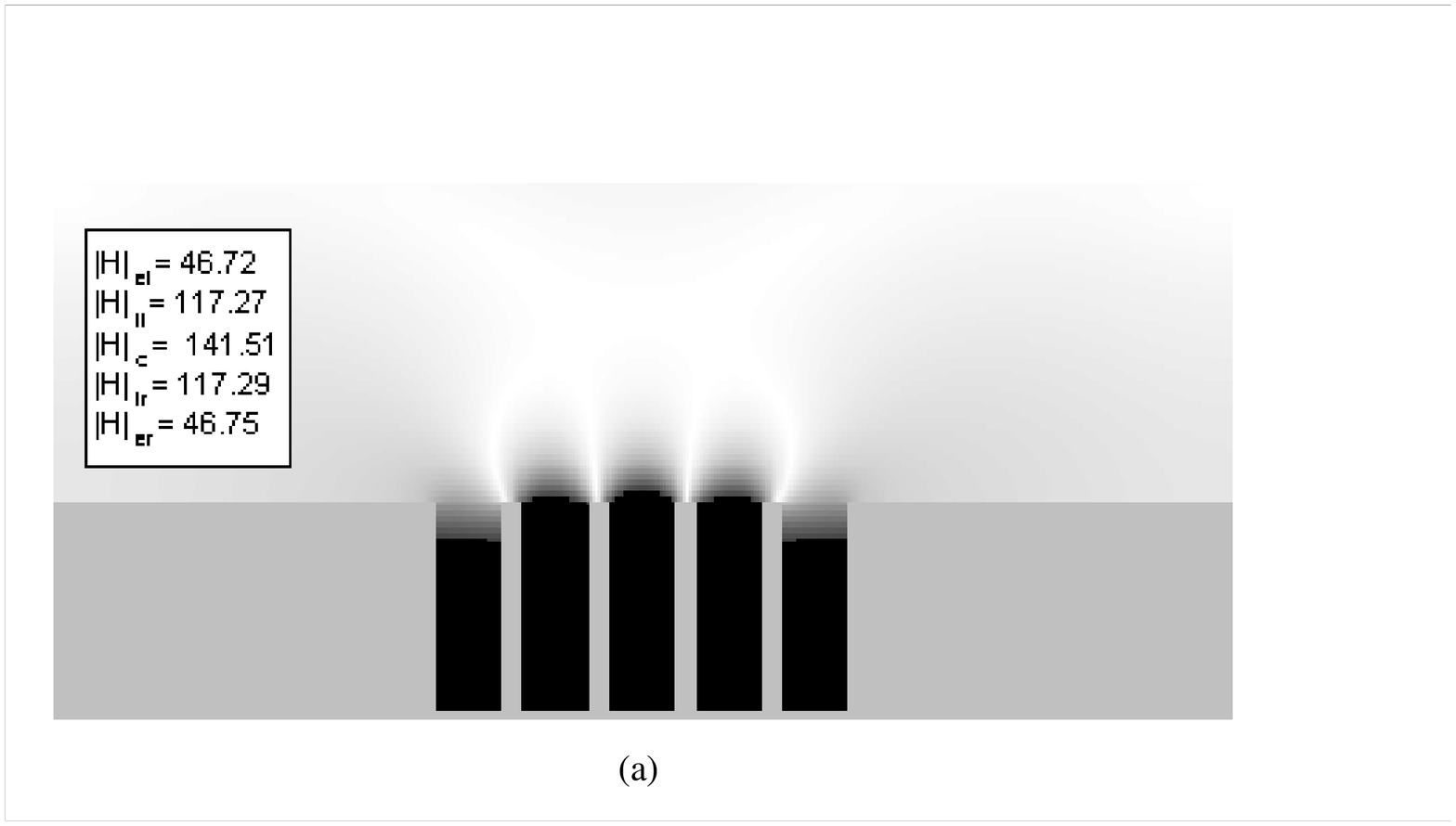} \\%\hspace{0.3cm}
\hspace{-1cm} \includegraphics[width=10cm]{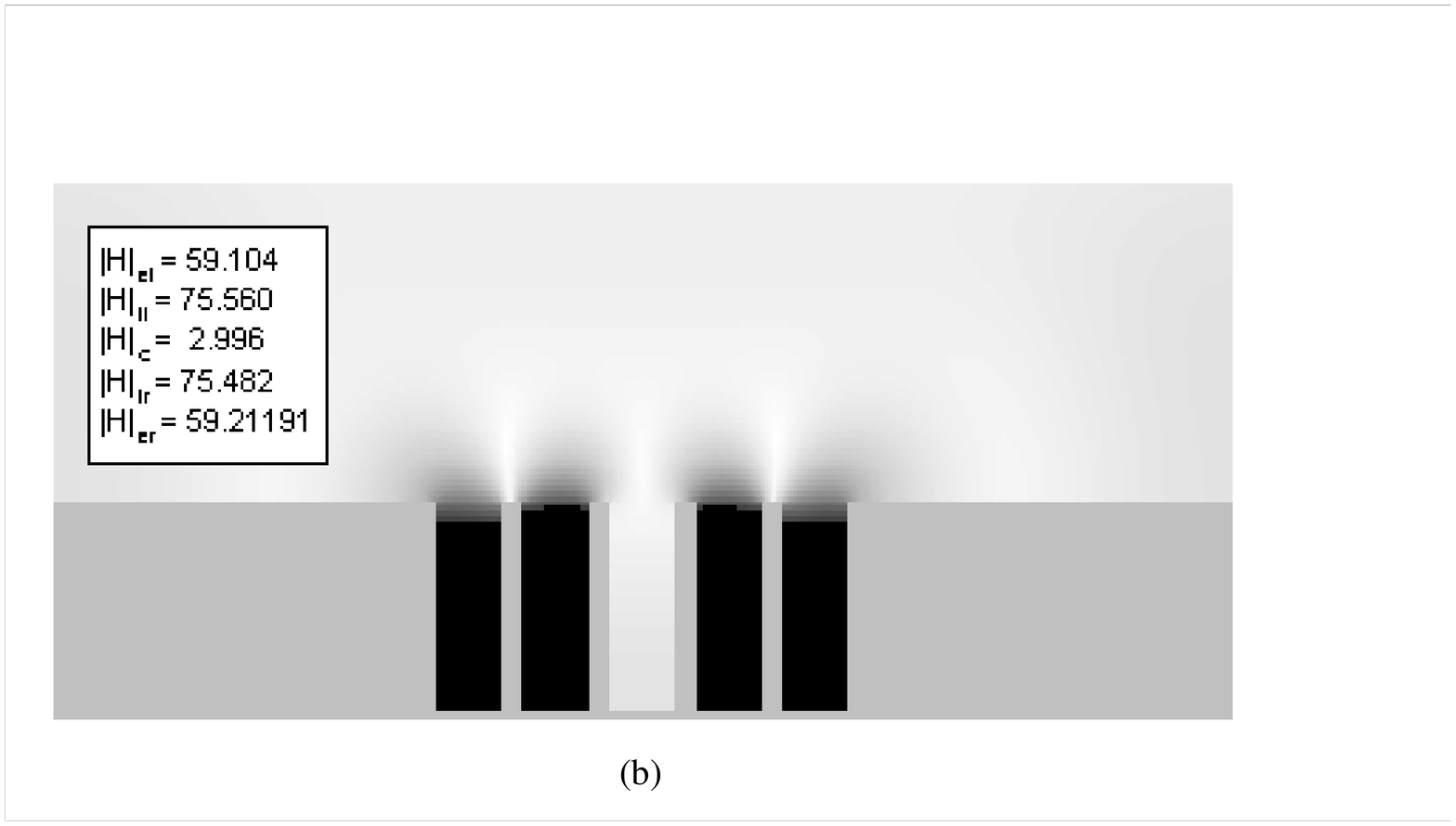} 
\hspace{-1.5cm} \includegraphics[width=10cm]{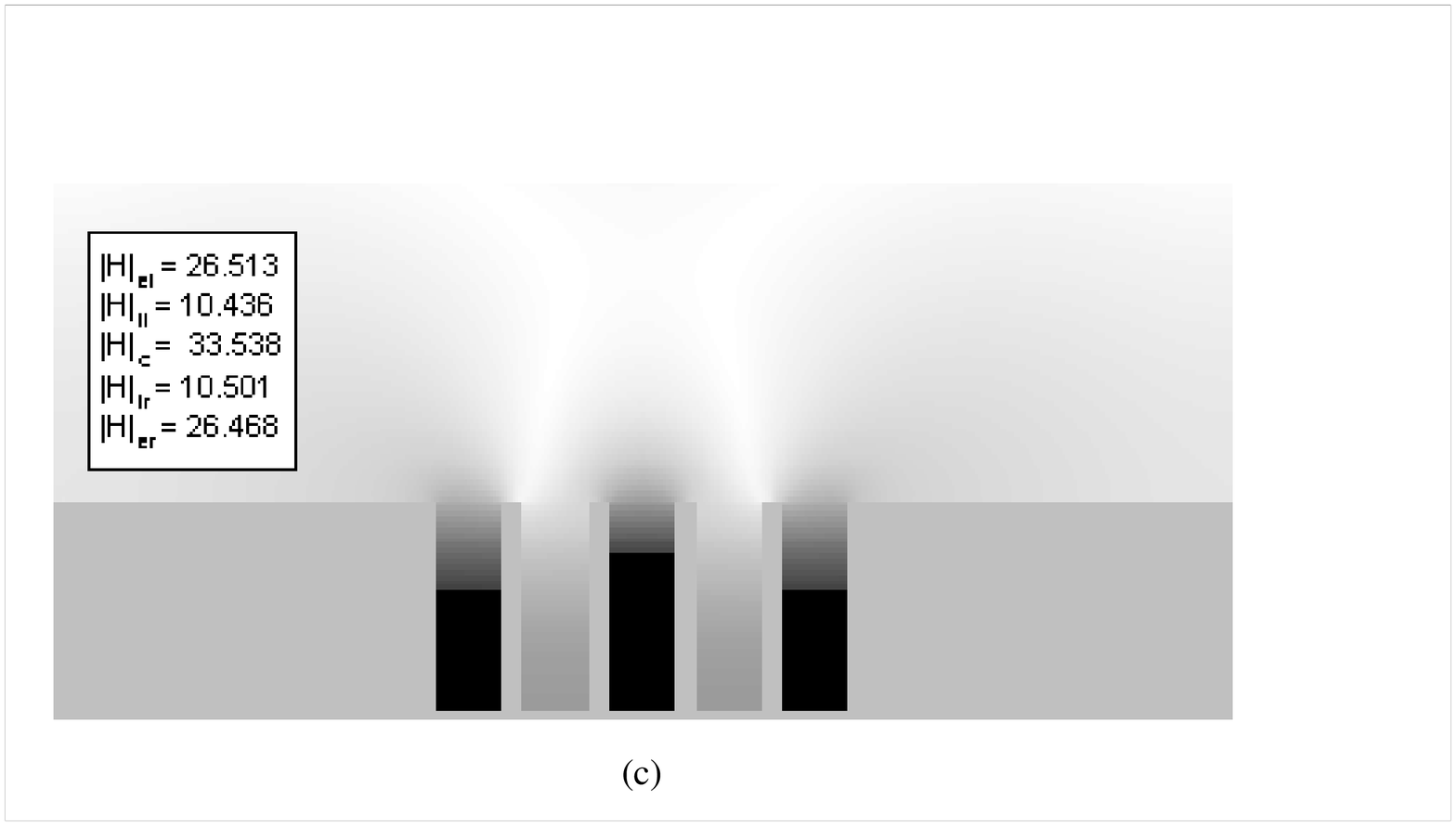} %\hspace{0.3cm}
\end{tabular}
\caption{Contour plots of the magnitude of the magnetic field in 
the vicinity of the grooves, for the resonant peaks of Fig. 6 (-1 Littrow mount): 
(a) $\theta_0=24.006^\circ$, $\lambda/h=4.475$; (b) $\theta_0=25.582^\circ$, $\lambda/h=4.575$; (c) $\theta_0=26.396^\circ$, $\lambda/h=4.890$.}
\end{figure}


\newpage
\begin{thebibliography}{99}

\bibitem{Agranovich} V. M. Agranovich and D. L. Mills, ``Surface polaritons'' (North
 Holland, Amsterdam, 1982)

\bibitem{Hessel} A. Hessel and A. A. Oliner, Appl. Opt. 4, 1275-1297 (1965).

\bibitem{Hutley} M. C. Hutley, Chapter 6 in {\em Diffraction gratings} (Academic Press,
London, 1982).

\bibitem{Valencia} C. I. Valencia and R. A. Depine, Opt. Commun. 159, 254-265 (1999).

\bibitem{diana13} D. C. Skigin and R. A. Depine, Phys. Rev. E 59, 3661-3668 (1999).

\bibitem{diana14} R. A. Depine and D. C. Skigin, Phys. Rev. E 61, 4479-4490 (2000).

\bibitem{Andrewartha} J. R. Andrewartha, J. R. Fox and I. J. Wilson, Optica Acta 26, 69-89 (1979).

\bibitem{Andrewartha3} J. R. Andrewartha, J. R. Fox and I. J. Wilson, Optica Acta 26, 
197-209 (1979).

\bibitem{Wirgin} A. Wirgin and A. A. Maradudin, Phys. Rev. B 31, 5573-5576 (1985).

\bibitem{Maradudin2} A. A. Maradudin, A. V. Shchegrov and T. A. Leskova, Opt. Commun. 135, 352-360 (1997).

\bibitem{Lopez-Rios} T. L\'opez-Rios, D. Mendoza, F. J. Garc\'{\i}a-Vidal,
J. S\'anchez-Dehesa and B. Pannetier, Phys. Rev. Lett. 81, 665-668 (1998).

\bibitem{fikioris} G. Fikioris, R. W. P. King and T. Wu, J. Appl. Phys. 68, 431-439 (1990).

\bibitem{Veremey2} V. V. Veremey and V. P. Shestopalov, Radio Science 26, 631-636 (1991).

\bibitem{Veremey1} V. V. Veremey, IEEE Antennas and Propagation Magazine 37, 16-27 (1995).

\bibitem{Veremey3} V. V. Veremey and R. Mittra, IEEE Trans. on Antennas Propag. 46, 4, 494-501
(1998).

\bibitem{dv1} D. C. Skigin, V. V. Veremey and R. Mittra, IEEE Trans. on Antennas Propag. 47, 
(2), 376-383 (1999).

\bibitem{bloch} A. Bloch, R. G. Medhurst, S. D. Pool and W. E. Knock, 
Proceedings of the IEE 48, 1164 (1960).


\bibitem{sad1} A. N. Fantino, S. I. Grosz and D. C. Skigin, Phys. Rev. E 64 (1), 016605 (2001).

\bibitem{sad2} S. I. Grosz, D. C. Skigin and A. N. Fantino, Phys. Rev. E 65 (5), 056619 (2002).

\bibitem{sad3} D. C. Skigin, A. N. Fantino and S. I. Grosz, J. Opt. A: Pure Appl. Opt. 5,
S129-S135 (2003).

\bibitem{Hibbins} A. P. Hibbins, J. R. Sambles and C. R. Lawrence, Appl. Phys. Lett. 80,
2410-2412 (2002).

\bibitem{Jovicevic} S. Jovicevic and S. Sesnic, J. Opt. Soc. Am. 62, 865-877 (1972).

\bibitem{DeSanto} J. A. DeSanto, J. Acoust. Soc. Am. 56, 1336-1341 (1974).

\bibitem{AndrewarthaDerrick1}  J. R. Andrewartha, G. H. Derrick and R. C.
McPhedran, Opt. Acta 28, 1177-1193 (1981).

\bibitem{Botten1} L. C. Botten, M. S. Craig, R. C. McPhedran, J. L. Adams and J. R. Andrewartha,
Opt. Acta 28, 1087-1102 (1981)

\bibitem{Botten2} L. C. Botten, M. S. Craig and R. C. McPhedran, Opt. Acta 28, 1103-1106 (1981).

\bibitem{Li} L. Li, J. Mod. Opt. 40, 553-573 (1993).

\bibitem{Fox} J. R. Fox, Opt. Acta 27, 289-305 (1980).

\bibitem{AndrewarthaDerrick2}  J. R. Andrewartha, G. H. Derrick and R. C. McPhedran, Opt. Acta 28,
1501-1516 (1981).

\bibitem{Moharam1} M. G. Moharam, J. Opt. Soc. Am. A 3, 1780-1787 (1986).

\bibitem{Li2} L. Li, J. Opt. Soc. Am. A 10, 2581-2591 (1993).

\bibitem{MerleElson} J. Merle Elson and P. Tran, J. Opt. Soc. Am. A 12, 1765-1771 (1995).

\bibitem{diana1} D. C. Skigin and R. A. Depine, Optik 94 (3), 114-122 (1993).


\end{thebibliography}
\end{document}